\documentclass[twocolumn,conference]{IEEEtran}
\usepackage[noadjust]{cite}
\IEEEoverridecommandlockouts
\usepackage{cite}
\usepackage{amsmath,amssymb,amsfonts}
\usepackage{algorithmic}
\usepackage{graphicx}
\usepackage{textcomp}
\usepackage{xcolor}
\usepackage{subfig}
\usepackage{ragged2e}
\usepackage{caption}
\usepackage{graphicx}
\usepackage{caption}
\makeatletter

\newcommand{\Rmnum}[1]{\expandafter\@slowromancap\romannumeral #1@}
\makeatother
\graphicspath{ {./picture/} }
\allowdisplaybreaks[4]
\def\BibTeX{{\rm B\kern-.05em{\sc i\kern-.025em b}\kern-.08em
    T\kern-.1667em\lower.7ex\hbox{E}\kern-.125emX}}

\title{Active Reconfigurable Intelligent Surface Empowered Synthetic Aperture Radar Imaging}

\author{\IEEEauthorblockN{Yifan Sun$^{\dag}$, Rang Liu$^{\ddag}$, Zhiping Lu$^{\star}$, Honghao Luo$^{\dag}$, Ming Li$^{\dag}$, and Qian Liu$^{\dag}$
\vspace{-0.0 cm} }

\IEEEauthorblockA{$^{\dag}$Dalian University of Technology, Dalian, Liaoning 116024, China}
\IEEEauthorblockA{$^{\ddag}$University of California, Irvine, CA 92697, USA}
\IEEEauthorblockA{$^{\star}$Beijing University of Posts and Telecommunications, Beijing 100876, China}
\IEEEauthorblockA{$^{\star}$State Key Laboratory of Wireless Mobile Communications (CICT), Beijing 100191, China\\ E-mail: \texttt{\{sunyifan, luohonghao\}@mail.dlut.edu.cn}, \texttt{rangl2@uci.edu},\\
\texttt{luzp@cict.com}, \texttt{\{mli, qianliu\}@dlut.edu.cn}}}

\begin{document}

\title{Active Reconfigurable Intelligent Surface Empowered Synthetic Aperture Radar Imaging\thanks{$^{\ast}$Corresponding author.}\thanks{This work is supported in part by the National Natural Science Foundation of China (Grant No.  62371090, 62071083,  and 62471086), in part by Liaoning Applied Basic Research Program (Grant No. 2023JH2/101300201 and 2023JH2/101700364), and in part by Dalian Science and Technology Innovation Project (Grant No. 2022JJ12GX014).}
}


\maketitle

\begin{abstract}
Synthetic Aperture Radar (SAR) utilizes the movement of the radar antenna over a specific area of interest to achieve higher spatial resolution imaging. In this paper, we aim to investigate the realization of SAR imaging for a stationary radar system with the assistance of active reconfigurable intelligent surface (ARIS) mounted on an unmanned aerial vehicle (UAV). As the UAV moves along the stationary trajectory, the ARIS can not only build a high-quality virtual line-of-sight (LoS) propagation path, but its mobility can also effectively create a much larger virtual aperture, which can be utilized to realize a SAR system. In this paper, we first present a range-Doppler (RD) imaging algorithm to obtain imaging results for the proposed ARIS-empowered SAR system. Then, to further improve the SAR imaging performance, we attempt to optimize the reflection coefficients of ARIS to maximize the signal-to-noise ratio (SNR) at the stationary radar receiver under the constraints of ARIS maximum power and amplification factor. An effective algorithm based on fractional programming (FP) and majorization minimization (MM) methods is developed to solve the resulting non-convex problem. Simulation results validate the effectiveness of ARIS-assisted SAR imaging and our proposed RD imaging and ARIS optimization algorithms.
\end{abstract}


\section{Introduction}
Synthetic aperture radar (SAR) stands out as a highly promising and versatile technique in the realm of remote sensing owing to its ability to generate high-resolution images and penetrate through various atmospheric conditions \cite{Curlander Wiley 1991}-\cite{Moreira GRSM 2013}.
In principle, SAR utilizes the movement of the radar antenna over a specific area of interest to achieve a larger virtual aperture and obtain higher-resolution imaging.
Consequently, SAR systems are often installed on mobile platforms like airplanes or spacecraft.
Exploring the use of widely deployed stationary radar systems to achieve SAR functionality has become an intriguing research topic.

Recently emerging reconfigurable intelligent surface (RIS) has drawn significant attention owing to its superior capability of manipulating radio environments by dynamically adjusting the phase-shift of the reflection element \cite{Wu C 2020}-\cite{Liu CST 2021}. Particularly when the line-of-sight (LoS) propagation path is obstructed, RIS can establish a virtual LoS path to overcome the blockage issue. Therefore, RIS has been envisioned as a prospective enabler and has been widely considered in the fields of wireless communications \cite{Pan JSTSP 2022}. Moreover, the deployment of RIS in radar systems also opens up opportunities for achieving more reliable and wider coverage sensing capability \cite{Luo TVT 2023}-\cite{Esmaeilbeig SPL 2022}.

Despite the many studies that have shown the benefits of passive RIS (PRIS) in communication systems, its major flaw, known as the ``multiplicative fading'' effect, has also been revealed. The loss of the reflection path created by the RIS is the result of multiplying the losses of the transmitter-RIS and RIS-target area propagation paths. This means that the RIS-empowered virtual LoS path has significant signal attenuation, which is even more severe for PRIS-aided radar systems since the signal may propagate via the PRIS twice. Active RIS (ARIS) is an innovative approach developed to efficiently address the multiplicative fading problem and effectively compensate for path loss by amplifying the reflected signal through the integration of reflection-type amplifiers into passive electromagnetic components \cite{Zhang TC 2023}-\cite{Zhu TCOM 2022}. Recent studies \cite{Zhu TVT 2023}-\cite{Rihan SPL 2022} have validated the benefits of ARIS and delved deeper into its application in radar systems.

On the other hand, the ARIS has features such as being lightweight, low cost, and small physical size, making it easily installable on an unmanned aerial vehicle (UAV). As the UAV moves along a stationary trajectory, the ARIS not only builds a high-quality virtual LoS propagation path, but its mobility can also effectively create a virtual aperture that is much longer than the physical antenna width. Therefore, one promising solution for achieving SAR imaging in stationary radar systems is to leverage the mobility afforded by a UAV-mounted ARIS.
This innovative approach enables conventional stationary radar systems to achieve high-resolution imaging of a certain area with the aid of a UAV-mounted ARIS.

Based on the above findings, in this paper, we aim to investigate how to utilize the UAV-mounted ARIS to realize and improve SAR imaging for a stationary radar station. Specifically, we first propose a range-Doppler (RD) imaging algorithm to obtain imaging results for the introduced ARIS-empowered SAR system. Then, to further improve the SAR imaging performance, we attempt to optimize the reflection coefficients of ARIS to direct the reflected radio signals toward the imaging region. To achieve this goal, the signal-to-noise ratio (SNR) at the stationary radar receiver is maximized under the constraints of ARIS maximum power and amplification factor. To tackle these non-convex optimization issues, we propose an efficient optimization method based on fractional programming (FP) \cite{Shen TSP 2018} and majorization minimization (MM) \cite{Sun TSP 2017}. Finally, simulation results validate the advancement of ARIS-assisted SAR imaging and the effectiveness of our proposed optimization algorithm.

\section{System Model}
\begin{figure}[!t]
  \centering
  \vspace{-0.0 cm}
  \includegraphics[width= 3 in]{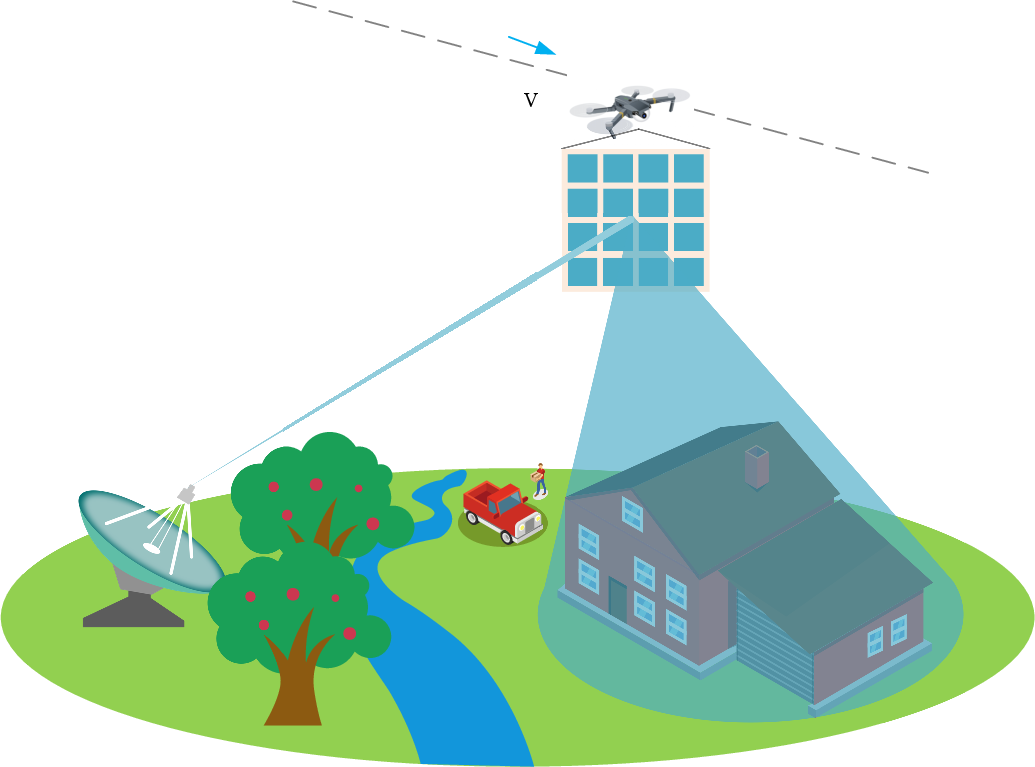}
  \caption{An ARIS-assisted SAR imaging system.}
  \vspace{-0.2 cm}
  \label{fig1}
\end{figure}

We consider an ARIS-assisted mono-static radar system as illustrated in Fig. \ref{fig1}.
The signals sent by the radar station are first amplified and reflected by an $M$-element ARIS toward the imaging area. Then, the echo signals will also propagate via the same path to the radar station for signal processing. As the ARIS is mounted on a UAV, the mobility of ARIS can form a virtual aperture to realize SAR imaging for a radar system with a stationary transceiver.
In specific, we assume that the transmission period is composed of $N$ slow time slots, each of which consists of $Q$ fast time slots. The duration of the slow and fast time slot is $\delta_t$ and $\delta_\tau$, respectively. Then, the transmitted signal at the $n$-th slow time slot and the $q$-th fast time slot can be expressed as
\begin{equation}
\label{eq: xnq}
x(n,q) = A_0\omega_{\rm r}(q){\rm exp}\left\{\jmath2\pi f_0q\delta_\tau+\jmath\pi k_{\rm r}(q\delta_\tau)^2\right\},
\end{equation}
where $A_0$, $f_0$, and $k_{\rm r}$ represent the amplitude, carrier frequency, and range modulation frequency, respectively. $\omega_{\rm r}(q)$ is the range rectangular envelope of the transmitted signal.

Denote $\mathbf h_{{\rm sr},n}$ and $\mathbf h_{{\rm {rt}},n}$ as the channels from the radar station to ARIS, and from ARIS to the imaging area respectively, which are assumed to follow Rician fading, e.g., $\mathbf h_{{\rm sr},n}$ can be expressed as
\begin{equation}
\label{2}
\mathbf h_{{\rm sr},n} = \alpha\sqrt{\frac{\kappa}{\kappa+1}}\mathbf{a}_{n}(\theta_{\rm sr}) +  \alpha\sqrt{\frac{1}{\kappa+1}}\mathbf{h_{\rm NLoS}},
\end{equation}
where $\alpha$ represents the distance-dependent path-loss, $\kappa$ is the Rician factor, $\mathbf{a}_{n}(\theta)$ represents the steering vector, $\theta_{\rm sr}$ is the direction of arrival (DoA), and $\mathbf{h_{\rm NLoS}}$ represents the non-LoS (NLoS) Rayleigh fading components. The reflection coefficient of the $m$-th element of the ARIS at the $n$-th slow time slot is defined as $\phi_{m,n}\triangleq a_{m,n} e^{\jmath\varphi_{m,n}}$, $m\in \{1, 2, \cdots, M\}$, where $a_{m,n}$ and $\varphi_{m,n}$ represent the amplitude and the phase-shift of the ARIS, respectively. The reflection matrix of the ARIS at the $n$-th time slot is defined as $\boldsymbol{\Phi}_n\triangleq{\rm diag}(\boldsymbol{\phi}_n)$ with $\boldsymbol{\phi}_n \triangleq \left[\phi_{1,n},\phi_{2,n},\cdots,\phi_{M,n}\right]^T$. Then, the transmitted signal that reaches the center point of the imaging area after being amplified and reflected by ARIS can be formulated as
\begin{equation}
\label{3}
r(n,q)= \mathbf{h}_{{\rm rt},n}^T \boldsymbol{\Phi}_n \left(\mathbf{h}_{{\rm sr},n}x\big(n,q-R_{n}/(c\delta_\tau)\big)+\mathbf{z}_0\right),
\end{equation}
where $R_{n}$ represents the propagation distance of the above-mentioned transmitted signal, and $c$ is the speed of light. The vector $\mathbf{z}_0\thicksim \mathcal{C}\mathcal{N}(0,\sigma_0^2\mathbf{I}_M)$ denotes the additive white Gaussian noise (AWGN) introduced by ARIS.

For simplicity, we divide the imaging area into an $N_{\rm a}\times N_{\rm r}$ grid shape. The scattering coefficient of each point target at the grid center is $g_{i,j}$, where $i\in \{1,2,\cdots, N_{\rm a}\}$ and $j\in \{1,2,\cdots, N_{\rm r}\}$. We set $g_{i,j} = 1$ if the point target exists in the grid and otherwise $g_{i,j} = 0$. Accordingly, after being reflected by the imaging area, the echo signal experiences ARIS reflection again and is finally collected by the radar station, the expression of which at the $n$-th slow time slot and $q$-th fast time slot can be written as
\begin{equation}
\small
\label{eq: ynq}
y(n,q)=\sum\nolimits_{i,j}\varpi_{\rm a}(n)g_{i,j}\big(h_{n}x_{i,j}(n,q)+\mathbf{h}_{0,n}^T\mathbf{z}_0\big)+\mathbf{h}_{1,n}^T\mathbf{z}_1+z,
\end{equation}
where we define the equivalent channel for the target return, the noise due to once reflection and twice reflection as
\begin{subequations}
\label{5a}
\begin{align}
\label{5b}
h_{n}&\triangleq \mathbf h_{{\rm sr},n}^T \boldsymbol{\Phi}_n\mathbf h_{ {\rm rt},n}\mathbf{h}_{{\rm rt},n}^T \boldsymbol{\Phi}_n\mathbf{h}_{{\rm sr},n},\\
\label{5c}
\mathbf{h}_{0,n}&\triangleq \boldsymbol{\Phi}_n\mathbf h_{ {\rm rt},n}\mathbf{h}_{{\rm rt},n}^T\boldsymbol{\Phi}_n\mathbf{h}_{{\rm sr},n},\\
\label{5d}
\mathbf{h}_{1,n}&\triangleq \boldsymbol{\Phi}_n\mathbf{h}_{{\rm sr},n}.
\end{align}
\end{subequations}
Besides, we define $x_{i,j}(n,q) \triangleq x\big(n,q-2R_{i,j,n}/(c\delta_\tau)\big)$ with the scalar $R_{i,j,n}$ denoting the propagation distance from the radar station to ARIS and then from ARIS to the $(i,j)$-th grid in the imaging area, $\varpi_{\rm a}(n) \triangleq \omega_{\rm a}(n-t_{i,j}/\delta_t)$ with $\omega_{\rm a}(n)$ representing the azimuth envelope of the received signal, and $t_{i,j}$ is the moment when the beam center passes through the target. $\mathbf{z}_1\thicksim \mathcal{C}\mathcal{N}(0,\sigma_0^2\mathbf{I}_M)$ represents the noise introduced by ARIS and $z\thicksim \mathcal{C}\mathcal{N}(0,\sigma^2)$ is AWGN.
Substituting the transmitted signal $x(n,q)$ in (\ref{eq: xnq}) into (\ref{eq: ynq}), the received signal can be re-arranged as
\begin{equation}
    y(n,q) =\!\! \sum\nolimits_{i,j}\!\!C_{i,j,n,q}{\rm exp}\left\{\jmath2\pi f_0r_{i,j,n,q}\!+\!\jmath\pi k_{\rm r}r_{i,j,n,q}^2\right\}
    +\widetilde{z},
\end{equation}
where we define the scalar
$C_{i,j,n,q}\triangleq g_{i,j}\varpi_{\rm a}(n)A_0\varpi_{\rm r}(q)h_{n}$ with $\varpi_{\rm r}(q)\triangleq\omega_{\rm r}(q-2R_{i,j,n}/({c}\delta_\tau))$, the scalar $r_{i,j,n,q}\triangleq q\delta_\tau-2R_{i,j,n}/c$, and the noise term $\widetilde{z}\triangleq \sum\nolimits_{i,j}\varpi_{\rm a}(n)g_{i,j}\mathbf{h}_{0,n}^T\mathbf{z}_0+\mathbf{h}_{1,n}^T\mathbf{z}_1 + z$.
To obtain the image of the sensing area, the received echo signal is further processed based on the RD imaging algorithm, which will be presented in the next section.

\section{Rang-Doppler Imaging Algorithm}
In this section, the RD imaging algorithm for the ARIS-empowered SAR system is presented to process the echo signal. The difference compared with conventional SAR systems lies in the involvement of ARIS, which establishes the sensing link by reflecting forward and backward signals. To make the algorithm development clearer, we will temporally ignore the noise term $\widetilde{z}$ in the following.

The received signals are first down-converted element-wise and collected by a two-dimensional matrix $\mathbf{Y} \in\mathbb{C}^{N\times Q}$, in which the $(n,q)$-th element is the down-converted version of $y(n,q)$, i.e., $\sum\nolimits_{i,j}{C}_{i,j,n,q}\exp\{-\jmath4\pi f_0 {R_{i,j,n}}/{c}+\jmath\pi k_{\rm r}\gamma_{i,j,n,q}^2\}$.
Then, range compression is employed to increase the range resolution while maintaining a longer pulse width to obtain sufficient signal power. Particularly, we define the range matched-filter as $\mathbf{h}_{\rm r}\triangleq[h_{\rm r}(1),h_{\rm r}(2),\cdots,h_{\rm r}(Q)]^T$, where $h_{\rm r}(q)\triangleq\exp\{-\jmath\pi k_{\rm r}(q\delta_\tau)^2\}$, and the signals after range compression as $\mathbf{Y}_\text{rc}\in\mathbb{C}^{N\times Q}
$. The $n$-th row of $\mathbf{Y}_\text{rc}$ is
\begin{equation}
\label{9}
\mathbf{Y}_\text{rc}(n,:)=\mathbf{Y}(n,:) \circledast \mathbf{h}_{\rm r}^T,
\end{equation}
in which $\circledast$ represents convolution, and the $q$-th element is calculated by $\mathbf{Y}_\text{rc}(n,q)=\sum\nolimits_{i,j}C_{i,j,n,q}p_{\rm r}(R_{i,j,n})\varphi_{\rm r}(R_{i,j,n})$ with the envelope of the compression pulse being defined as $p_{\rm r}(R)\triangleq{\rm sinc}\left(q\delta_\tau-{2R}/{c}\right)$ and the exponential component being $\varphi_{\rm r}(R)\triangleq \exp\{-\jmath4\pi f_0{R}/{c}\}$.

To implement ARIS-assisted SAR imaging using the echo signals received at the radar station, an appropriate RD imaging algorithm should be developed. A crucial step in this process is to remove the temporal delay of the echo signal that occurs between the radar station and ARIS since only the instantaneous range distance from ARIS to the imaging area at each slow time slot is crucial for subsequent imaging processing. Specifically, the distance between the radar station and imaging area equals to $R_{i,j,n} = R_{{\rm rt}, i,j,n}+R_{{\rm sr},n}$, where $R_{{\rm rt}, i,j,n}$ is the distance between ARIS and the $(i,j)$-th grid of the imaging area, and $R_{{\rm sr},n}$ is the distance between the radar station and ARIS.
By using the impulse sequence $\boldsymbol{{\delta}}\triangleq[{\delta}(1),{\delta}(2),\cdots,{\delta}(Q)]^T$, where ${\delta}(q)=1$ if $q=\lfloor R_{{\rm sr},n}/\delta_\tau \rfloor$ and elsewhere ${\delta}(q)=0$, the temporal delay can be removed and the output is
\begin{equation}
    \widetilde{\mathbf{Y}}(n,:) = (\mathbf{Y}_\text{rc}(n,:)\circledast \boldsymbol{{\delta}})\varphi_{\rm r}(-R_{{\rm sr},n}),
\end{equation}
where the $q$-th element can be expressed as $\sum\nolimits_{i,j}C_{i,j,n,q}p_{\rm r}(R_{{\rm rt},i,j,n})\varphi_{\rm r}(R_{{\rm rt},i,j,n})$. This strategic modification improves the adaptability of conventional RD imaging techniques in this novel scenario involving mobile ARIS, facilitating the radar station-based SAR imaging functionality.

Since the distance between the target and the ARIS will continuously change with the mobility of the ARIS, we perform the range migration correction (RCMC) on the signal. In specific, we first perform the azimuth Fourier transform. For brevity, we assume $R_{{\rm rt}, i,j,n}\triangleq R_{{\rm 0}, i,j,n}+{v^2(n\delta_t)^2}/(2R_{{\rm 0}, i,j,n})\approx R_{{\rm 0}, i,j,n}+{v^2(n\delta_t)^2}/(2R_{n})$, where $R_{{\rm 0}, i,j,n}$ is the slant range corresponding to arbitrary targets' zero Doppler time $t_{i,j}$, namely, the moment when ARIS is closest to the point target, and $v$ is the velocity of the ARIS. According to the principle of stationary phase (POSP) \cite{Cumming Artech House 2005}, the time-frequency relationship used for FFT is given by
\begin{equation}
\label{11}
f=-k_{\rm a}n\delta_t,
\end{equation}
where $k_{\rm a}\triangleq 2v^2/(\lambda R_{n})$ is the azimuth modulation frequency.
The output after employing FFT along each column of $\widetilde{\mathbf{Y}}$ is
\begin{equation}
\label{12}
\widehat{\mathbf{Y}}(:,q)={\rm FFT}\{\widetilde{\mathbf{Y}}(:,q)\},~\forall q.
\end{equation}
The $(n,q)$-th element of $\widehat{\mathbf{Y}}$ is calculated by $\widehat{\mathbf{Y}}(n,q)=\sum\nolimits_{i,j}{C}_{i,j,n,q}p_{\rm r}(\widehat{R}_{{\rm f},i,j,n})\omega_{\rm a}(n-f_{i,j}/\delta_f)\varphi_{\rm r}(R_{{\rm 0},i,j,n})\exp\left\{\jmath\pi (n\delta_{f})^2/k_{\rm a}\right\}$, where we define $\widehat{R}_{{\rm f},i,j,n}\triangleq R_{{\rm 0},i,j,n}+\lambda^2R_{n}(n\delta_{f})^2/(8v^2)$, $\delta_{f}\triangleq 1/(N\delta_t)$, and $\omega_{\rm a}(n-f_{i,j}/\delta_f)$ is the Fourier transform of $\omega_{\rm a}(n-t_{i,j}/\delta_t)$. Afterward, utilizing the RCMC algorithm with the sinc interpolation, the matrix $\widehat{\mathbf{Y}}$ is further processed, and the $(n,q)$-th element of which is given by $\widehat{\mathbf{Y}}_\text{s}(n,q)\triangleq\sum\nolimits_{i,j}{C}_{i,j,n,p}p_{\rm r}(R_{{\rm 0}, i,j,n})\varphi_{\rm r}(R_{{\rm 0}, i,j,n})\omega_{\rm a}(n-f_{i,j}/\delta_f)\exp\left\{\jmath\pi (n\delta_{f})^2/k_{\rm a}\right\}$.
Then, an azimuth matched filter $\mathbf{h}_{\rm a}\triangleq[h_{\rm a}(1), h_{\rm a}(2), \cdots, h_{\rm a}(N)]^T$ with $h_{\rm a}(n)\triangleq\exp\{-\jmath\pi {(n\delta_{f})^2}/{k_{\rm a}}\}$ is employed for each column of $\widehat{\mathbf{Y}}_\text{s}$. Finally, an inverse FFT is used to obtain the imaging result $\overline{\mathbf{Y}}$. The $q$-th column of $\overline{\mathbf{Y}}$ can be calculated as
\begin{equation}
\label{14}
\overline{\mathbf{Y}}(:,q)= {\rm IFFT}\big\{\widehat{\mathbf{Y}}_\text{s}(:,q)\odot \mathbf{h}_{\rm a}\big\},
\end{equation}
of which $\odot$ represents Hadamard product, and the $n$-th element of $\overline{\mathbf{Y}}(:,q)$ can be expressed as
\begin{equation}\label{eq: output}
\overline{\mathbf{Y}}(n,q)\!\!\triangleq\!\!\sum\nolimits_{i,j}\!{C}_{i,j,n,p}p_{\rm r}(R_{{\rm 0},i,j,n})p_{\rm a}(t_{i,j})\varphi_{\rm r}(R_{{\rm 0},i,j,n})\varphi_{\rm a}(f_{i,j}).
\end{equation}
In (\ref{eq: output}), $p_{\rm a}(t)$ is the azimuth sinc($\cdot$) function, which is denoted as $p_{\rm a}(t)\triangleq {\rm sinc}(n\delta_t-t_{i,j})$, and $\varphi_{\rm a}(f_{i,j})\triangleq\exp\left\{\jmath2\pi f_{i,j}n\delta_t\right\}$.
With the output $\overline{\mathbf{Y}}$, we can easily obtain the imaging picture.
It is straightforward that a higher amplitude of $\overline{\mathbf{Y}}(n,q)$ allows for better noise immunity and imaging results.

\section{ARIS Design Algorithm}
The quality of SAR imaging is significantly influenced by the echo signal SNR at the radar receiver. In order to achieve better SAR imaging, in this section we aim to design the phase-shift and amplitude of ARIS to focus the reflected signal toward the imaging area at each slow time, thereby providing higher SNR at the receiver.

By defining $P_{s}\triangleq |x(n,q)|^2$, the expression of the SNR can be calculated as
\begin{equation}
\label{15}
{\rm SNR}_n=\frac{P_{s}|h_{n}|^2}{\sigma^2+\sigma_0^2\mathbf{h}_{0,n}^T\mathbf{h}_{0,n}^* + \sigma_1^2\mathbf{h}_{1,n}^T\mathbf{h}_{1,n}^*}.
\end{equation}
Then, the ARIS design problem can be formulated as
\begin{subequations}
\label{16}
\begin{align}
\label{16a}
\max_{\boldsymbol{\phi}_n} \quad & {\rm SNR}_n \\
\label{16b}
\text{s.t.} \quad & P(\boldsymbol{\phi}_n) \leq P_{\rm ARIS},\\
\label{16c}
& a_{m,n} \leq a_{\rm max}, \quad \forall m,~ n,
\end{align}
\end{subequations}
where $P_{\rm ARIS}$ is the power constraint of the ARIS, $a_{\rm max}$ is the maximum amplification gain of $a_{m,n}$, and the power consumed at the ARIS is given by
\begin{equation}\begin{aligned}
   P(\boldsymbol{\phi}_n) &\triangleq P_{\rm s}\|\boldsymbol{\Phi}_n\mathbf{h}_{{\rm sr},n}\|_2^2+\sigma_0^2\|\boldsymbol{\Phi}_n\mathbf{h}_{{\rm rt},n}\mathbf{h}_{{\rm rt},n}^T\boldsymbol{\Phi}_n\|_F^2  \\
   &+P_{\rm s}\|\boldsymbol{\Phi}_n\mathbf{h}_{{\rm rt},n}\mathbf{h}_{{\rm rt},n}^T\boldsymbol{\Phi}_n\mathbf{h}_{{\rm sr},n}\|_2^2+(\sigma_0^2\!+\!\sigma_1^2)\|\boldsymbol{\Phi}_n\|_F^2.
\end{aligned}\end{equation}
In the following, we propose an efficient alternating optimization-based algorithm with the assistance of FP and MM methods.

We find that problem \eqref{16} is difficult to solve due to the complicated objective function \eqref{16a} and the non-convex constraint \eqref{16b}.
To address the complicated fractional objective function corresponding to the variable $\boldsymbol{\phi}_n$, we utilized the FP algorithm \cite{Shen TSP 2018} to transform \eqref{16a} into a polynomial form by introducing auxiliary variable $l_{n}$. Therefore, the original objective function is transformed into
\begin{subequations}
\label{17}
\begin{align}
\label{17a}
&P_s|h_{n}|^2-l_{n}(\sigma^2+\sigma_0^2\mathbf{h}_{0,,n}^T\mathbf{h}_{0,n}^* + \sigma_1^2\mathbf{h}_{1,n}^T\mathbf{h}_{1,n}^*)\\
\label{17b}
&= {\rm vec}^H\{\boldsymbol{\phi}_n\boldsymbol{\phi}_n^T\}\mathbf{D}_n{\rm vec}\{\boldsymbol{\phi}_n\boldsymbol{\phi}_n^T\}-\boldsymbol{\phi}_n^H\mathbf{V}_n\boldsymbol{\phi}_n-\widehat{c}_n ,
\end{align}
\end{subequations}
where for brevity, we define
\begin{subequations}
\label{18}
\begin{align}
 \mathbf{D}_n &\triangleq  P_{\rm s}\mathbf{A}_{n}\otimes\mathbf{A}_{n}-l_{n}\sigma_0^2(\mathbf{B}_{n}\otimes\mathbf{A}_{n}),\\
 \label{18d}
 \mathbf{V}_n &\triangleq l_{n}\sigma^2{\rm diag}\{\mathbf{h}_{{\rm sr},n}^*\}{\rm diag}\{\mathbf{h}_{{\rm sr},n}\},\\
 \label{18e}
 \mathbf{A}_{n}&\triangleq {\rm diag}\{\mathbf{h}_{{\rm rt},n}^*\}\mathbf{h}_{{\rm sr},n}\mathbf{h}_{{\rm sr},n}^T{\rm diag}\{\mathbf{h}_{{\rm rt},n}\},\\
 \label{18f}
 \mathbf{B}_{n}&\triangleq{\rm diag}\{\mathbf{h}_{{\rm rt},n}^*\}{\rm diag}\{\mathbf{h}_{{\rm rt},n}\},\\
 \label{18g}
\widehat{c}_n&\triangleq l_{n}\sigma^2.
 \end{align}
\end{subequations}
And the optimal value of the auxiliary variable $l_{n}$ after each iteration is
\begin{equation}
\label{19}
l^\star_{n} = \frac{P_s|h_{n}|^2}{\sigma^2+\sigma_0^2\mathbf{h}_{0,,n}^T\mathbf{h}_{0,n}^* + \sigma_1^2\mathbf{h}_{1,n}^T\mathbf{h}_{1,n}^*},~ \forall n.
\end{equation}

However, the first term of \eqref{17b} is a quartic form, which is still complicated to solve directly. Therefore, we employ the MM method to find a convex surrogate function in each iteration \cite{Sun TSP 2017}. To be specific, by using the first-order Taylor expansion, a lower bound of \eqref{17b} can be expressed as
\begin{equation}
\label{20}
{\rm vec}^H\{\boldsymbol{\phi}_n\boldsymbol{\phi}_n^T\}\mathbf{D}_n{\rm vec}\{\boldsymbol{\phi}_n\boldsymbol{\phi}_n^T\}\geq \Re\{(\mathbf f_n^k)^H {\rm vec}\{\boldsymbol{\phi}_n\boldsymbol{\phi}_n^T\}\} +c_0,
\end{equation}
where we define the vector $\mathbf{f}_n^k\triangleq 2 \mathbf D_n{\rm vec}\{\boldsymbol{\phi}_n^k(\boldsymbol{\phi}_n^k)^T\}$, the scalar $c_0\triangleq{\rm vec}^H\{\boldsymbol{\phi}_n^k(\boldsymbol{\phi}_n^k)^T\}\mathbf D_n{\rm vec}\{\boldsymbol{\phi}_n^k(\boldsymbol{\phi}_n^k)^T\}$, and $\boldsymbol{\phi}_n^k$ is the value of $\boldsymbol{\phi}_n$ from the previous iteration. Substituting \eqref{16a} into \eqref{17b} and \eqref{20} and defining $\mathbf{f}_n^k\triangleq {\rm vec}\{\mathbf{F}_n^k\}$, the original problem function can be reformulated as
\begin{subequations}
\label{21}
\begin{align}
\label{21a}
\max\limits_{\boldsymbol{\phi}_n}\quad &\Re\{\boldsymbol{\phi}^H_n\mathbf{F}_n^k\boldsymbol{\phi}_n^*\}-\boldsymbol{\phi}_n^H\mathbf{V}_n\boldsymbol{\phi}_n\\
\label{21b}
\text{s.t.} \quad & P(\boldsymbol{\phi}_n) \leq P_{\rm ARIS}, \\
& a_{m,n} \leq a_{\rm max},~~\forall m,~n.
\end{align}
\end{subequations}

Since the first term of \eqref{21a} is still non-convex with respect to ${\boldsymbol{\phi}_n}$, further conversion of quadratic terms to one-order term is required. Firstly, in order to convert it into a real-valued form, we perform the following transformation
\begin{equation}
\label{22}
\Re\{\boldsymbol{\phi}^H_n\mathbf{F}_n^k\boldsymbol{\phi}_n^*\}=\overline{\boldsymbol{\phi}}_n\overline{\mathbf{F}}_n^k\overline{\boldsymbol{\phi}}_n,
\end{equation}
where we define the vector $\overline{\boldsymbol{\phi}}_n \triangleq \begin{bmatrix} \Re\{{\boldsymbol \phi}_n^T\}, \Im\{\boldsymbol{\phi}_n^T\} \end{bmatrix}^T$ and the matrix $\overline{\mathbf{F}}_n^k \triangleq \begin{bmatrix} \Re\{\mathbf{F}_n^k\} &\Im\{\mathbf{F}_n^k\} \\ \Im\{\mathbf{F}_n^k\} & -\Re\{\mathbf{F}_n^k\}\end{bmatrix}$. The first term of \eqref{21a} then can be processed through first-order Taylor expansion
\begin{equation}
\label{23}
\begin{aligned}
\overline{\boldsymbol{\phi}}_n^T\overline{\mathbf{F}}_n^k\overline{\boldsymbol{\phi}}_n
\geq&(\overline{\boldsymbol{\phi}}_n^k)^T\overline{\mathbf{F}}_n^k\overline{\boldsymbol{\phi}}_n^k+(\overline{\boldsymbol{\phi}}_n^k)^T(\overline{\mathbf{F}}_n^k+(\overline{\mathbf{F}}_n^k)^T)(\overline{\boldsymbol{\phi}}_n-\overline{\boldsymbol{\phi}}_n^k)\\
=&\Re\{\boldsymbol{\phi}^H_n \mathbf{\overline{f}}_n^k\}+c_1,
\end{aligned}
\end{equation}
where we define the scalar $c_1 \triangleq -(\overline{\boldsymbol{\phi}}_n^k)^T\overline{\mathbf{F}}_n^k\overline{\boldsymbol{\phi}}_n^k$, the vector $\mathbf{\overline{f}}_n^k \triangleq \mathbf{U}(\overline{\mathbf{F}}_n^k+(\overline{\mathbf{F}}_n^k)^T)\overline{\boldsymbol{\phi}}_n^k$, and the matrix $\mathbf{U} \triangleq \begin{bmatrix}\mathbf{I}_M & \jmath\mathbf{I}_M \end{bmatrix}$.

Similarly, the first constraint of the original problem \eqref{16b} can be simplified as
\begin{equation}
\label{24}
P({\boldsymbol \phi}_n)={\rm vec}^H\{\boldsymbol{\phi}_n\boldsymbol{\phi}_n^T\}\mathbf{H}_n{\rm vec}\{\boldsymbol{\phi}_n\boldsymbol{\phi}_n^T\}+\boldsymbol{\phi}_n^H\mathbf{G}_n\boldsymbol{\phi}_n,
\end{equation}
where for brevity we define
\begin{subequations}
\label{25}
\begin{align}
\label{25a}
\mathbf{H}_n&\triangleq P_{\rm s}\mathbf{A}_{n}\otimes\mathbf{B}_{n}+\sigma_0^2\mathbf{B}_{n}\otimes\mathbf{B}_{n},\\
\label{25b}
\mathbf{G}_n&\triangleq P_{\rm s}{\rm diag}\{\mathbf{h}_{{\rm sr},n}^*\}{\rm diag}\{\mathbf{h}_{{\rm sr},n}\}+(\sigma_0^2+\sigma_1^2)\mathbf{I}_M.
\end{align}
\end{subequations}

To deal with the quartic term ${\rm vec}^H\{\boldsymbol{\phi}_n\boldsymbol{\phi}_n^T\}\mathbf{H}_n{\rm vec}\{\boldsymbol{\phi}_n\boldsymbol{\phi}_n^T\}$ in \eqref{24}, we utilize the MM method to find an upper bound of it as
\begin{subequations}
\label{26}
\begin{align}
\label{26a}
{\rm vec}^H\{\boldsymbol{\phi}_n\boldsymbol{\phi}_n^T\}\mathbf{H}_n{\rm vec}\{\boldsymbol{\phi}_n\boldsymbol{\phi}_n^T\}\leq& \Re\{{\rm vec}^H\{\boldsymbol{\phi}_n\boldsymbol{\phi}_n^T\}\mathbf{q}_n^k\}+c_2\\
\label{26b}
=&\overline{\boldsymbol{\phi}}_n^T\overline{\mathbf{Q}}_n^k\overline{\boldsymbol{\phi}}_n+c_2,
\end{align}
\end{subequations}
where we define $\mathbf{q}_n^k\triangleq 2\left(\mathbf{H}_n-\lambda_1\mathbf{I}_{M^2}\right){\rm vec}\{\boldsymbol{\phi}_n^k(\boldsymbol{\phi}_n^k)^T\}\triangleq {\rm vec}\{\mathbf{Q}_n^k\}$, $\lambda_1$ is the maximum eigenvalue of $\mathbf{H}_n$, $c_2\triangleq \lambda_1M^2a_{\rm max}^4+{\rm vec}^H\{\boldsymbol{\phi}_n^k(\boldsymbol{\phi}_n^k)^T\}(\lambda_1\mathbf{I}_{M^2}-\mathbf{H}_n){\rm vec}\{\boldsymbol{\phi}_n^k(\boldsymbol{\phi}_n^k)^T\}$, and $\overline{\mathbf{Q}}_n^k \triangleq \begin{bmatrix} \Re\{\mathbf{Q}_n^k\} &\Im\{\mathbf{Q}_n^k\} \\ \Im\{\mathbf{Q}_n^k\} & -\Re\{\mathbf{Q}_n^k\}\end{bmatrix}$.

However, the term $\overline{\boldsymbol{\phi}}_n\overline{\mathbf{Q}}_n^k\overline{\boldsymbol{\phi}}_n$ is still non-convex with respect to $\boldsymbol{\phi}$. Thus, we simplify the first term of the above equation using second-order Taylor expansion
\begin{subequations}
\label{27}
\begin{align}
\label{27a}
\overline{\boldsymbol{\phi}}_n\overline{\mathbf{Q}}_n^k\overline{\boldsymbol{\phi}}_n
\leq&\frac{\lambda_2}{2}(\overline{\boldsymbol{\phi}}_n-\overline{\boldsymbol{\phi}}_n^k)^T(\overline{\boldsymbol{\phi}}_n-\overline{\boldsymbol{\phi}}_n^k)\\
+(\overline{\boldsymbol{\phi}}_n^k)^T(\overline{\mathbf{Q}}_n^k&+(\overline{\mathbf{Q}}_n^k)^T)(\overline{\boldsymbol{\phi}}_n-\overline{\boldsymbol{\phi}}_n^k)+(\overline{\boldsymbol{\phi}}_n^k)^T\overline{\mathbf{Q}}_n^k\overline{\boldsymbol{\phi}}_n^k \notag\\
\label{27b}
=&\frac{\lambda_2}{2}\boldsymbol{\phi}^H_n\boldsymbol{\phi}_n+\Re\{\boldsymbol{\phi}_n^H\mathbf{\hat{p}}_n^k\}+c_3,
\end{align}
\end{subequations}
where $\mathbf{\widehat{p}}_n^k\triangleq \mathbf{U}(\overline{\mathbf{Q}}_n^k+(\overline{\mathbf{Q}}_n^k)^T)\overline{\boldsymbol{\phi}}_n^k$, $\lambda_2$ is the maximum eigenvalue of $(\overline{\mathbf{Q}}_n^k+(\overline{\mathbf{Q}}_n^k)^T)$, and $c_3=-\lambda_2(\overline{\boldsymbol{\phi}}_n^k)^T\overline{\boldsymbol{\phi}}_n^k/2$.

Finally, the original problem \eqref{16} can be reformulated as
\begin{subequations}
\label{28}
\begin{align}
\label{28a}
\max\limits_{\boldsymbol{\phi}_n}\quad &-\boldsymbol{\phi}_n^H\mathbf{V}_n\boldsymbol{\phi}_n+\Re\{\boldsymbol{\phi}_n^H\mathbf{\overline{f}}_n^k\}\\
\label{28b}
{\rm s.t.}\quad&\boldsymbol{\phi}_n^H\mathbf{K}_n\boldsymbol{\phi}_n +\Re\{\boldsymbol{\phi}_n^H\mathbf{\hat{p}}_n^k\}+c_2+c_3 \leq P_{\rm ARIS},\\
\label{28c}
&a_m \leq a_{\rm max}, \quad\forall m,
\end{align}
\end{subequations}
where $\mathbf{K}_n\triangleq \mathbf{G}_n+{\lambda_2}/{2}\mathbf{I}_M$. It is found that the above problem is convex and can be solved by CVX.

\section{Simulation Results}

\begin{figure}
  \centering
  \includegraphics[width= 0.75\linewidth ]{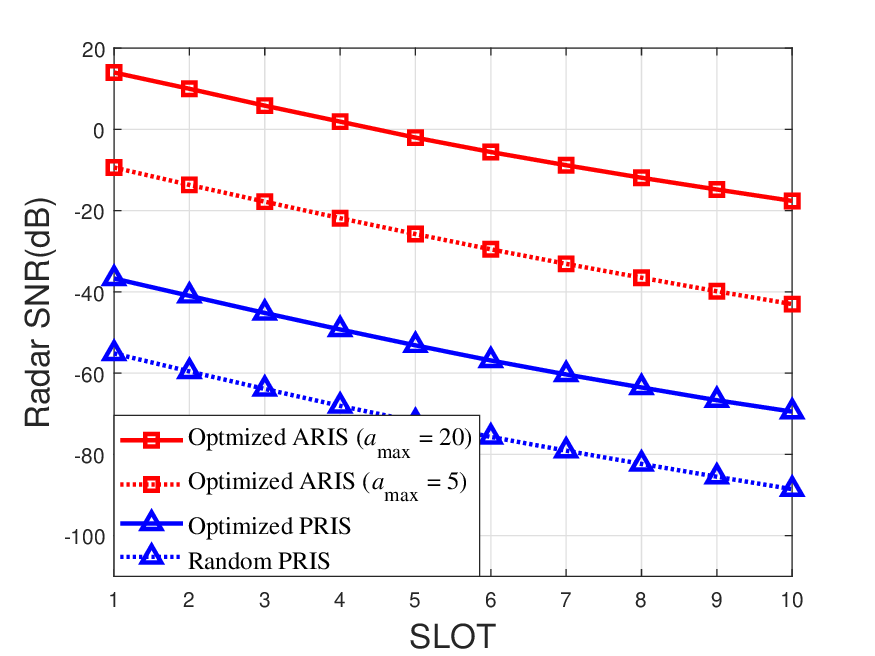}
  \caption{Radar SNR versus time slot.}\vspace{-0.2 cm}
  \label{fig2}
\end{figure}
\begin{figure}
  \centering
  \includegraphics[width= 0.75\linewidth ]{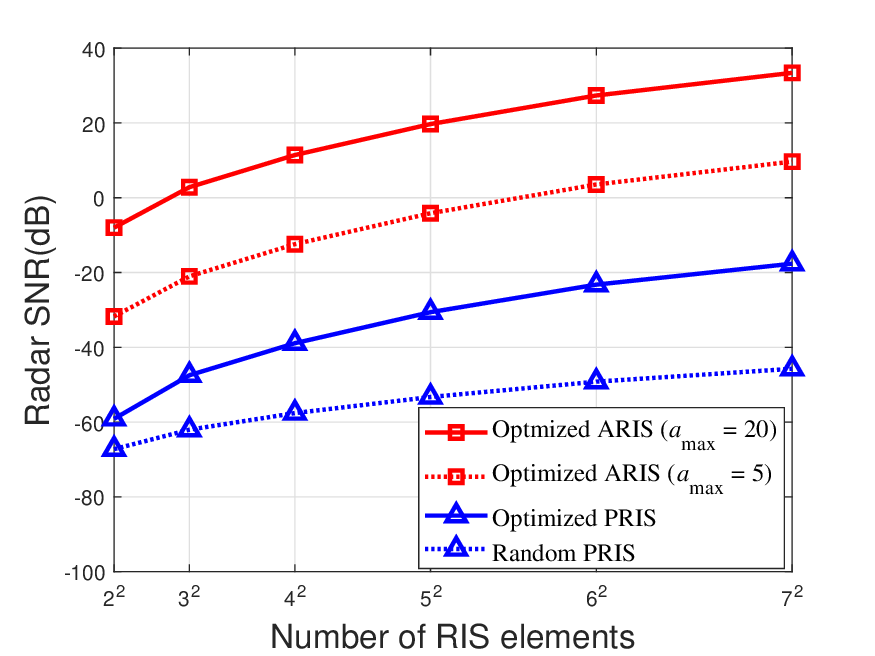}
  \caption{Radar SNR versus numbers of RIS elements.}\vspace{-0.2 cm}
  \label{fig3}
\end{figure}

In this section, we provide simulation results to demonstrate the efficiency and advancement of our proposed algorithm. We assume that the size of the imaging area is $30\times30 \rm m^2$, the closet distance between ARIS and the center of the imaging area is $300 \rm m$, the initial distance between the radar station and the ARIS is $5\rm m$, and the velocity of the UAV is 30m/s with the height being 20m. The center frequency, bandwidth, and pulse repetition frequency of the transmitted signal are 6GHz, 300MHz, and 720Hz, respectively. Besides, we adopt the Rician fading model,  of which the path-loss exponents for the link between the radar station and ARIS and the link between ARIS and a certain grid of the target area are 2.2 and 2.2 respectively. Also, the Rician factor for the above two links is set as $\kappa=3{\rm dB }$. The additive Gaussian noise power is set as $\sigma=\sigma_0=\sigma_1=-80{\rm dBm}$.

We first present the radar SNR versus different time slots in Fig. \ref{fig2}. The PRIS schemes are also included for better comparison. In the ARIS scheme, the transmit power of radar is $P_{\rm s}=85{\rm W}$, and the constraint power of ARIS is $P_{\rm ARIS}=15{\rm W}$. For fairness,  we set $P_{\rm radar} + P_{\rm ARIS}$ as the transmit power in the PRIS case. Compared with PRIS, the SNR gain of ARIS is much higher since ARIS enhances the incident signal by amplifying it, thereby mitigating the effects of multiplicative fading. Over time, the distance between ARIS and radar station increases, leading to a decrease in the SNR.

\begin{figure}
  \centering
  \includegraphics[width= 0.75\linewidth ]{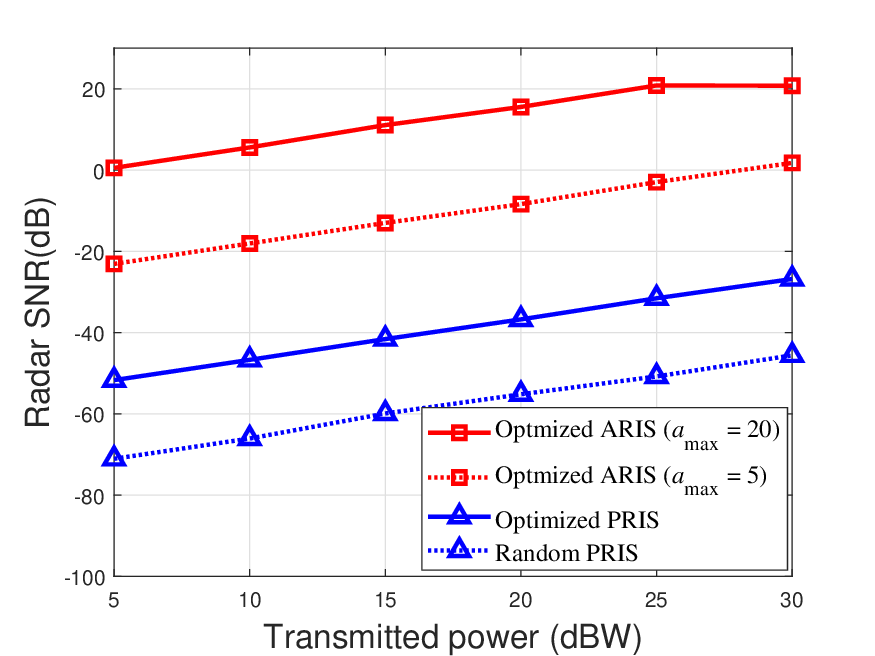}
  \caption{Radar SNR versus transmitted power.}\vspace{-0.2 cm}
  \label{fig4}
\end{figure}

Afterward, we investigate the impact of the number of ARIS elements on the radar SNR in Fig. \ref{fig3}. As the number of ARIS elements increases, the radar SNR gradually increases in both ARIS and PRIS. Besides, the performance gap between the optimized PRIS scheme and the random PRIS scheme is widening, whereas the gap between the optimized ARIS with $a_{\rm max}=20$ scheme and the optimized ARIS with $a_{\rm max}=5$ scheme remains the same.

\begin{figure*}
\centering
\includegraphics[width= 15cm, height=3.75cm]{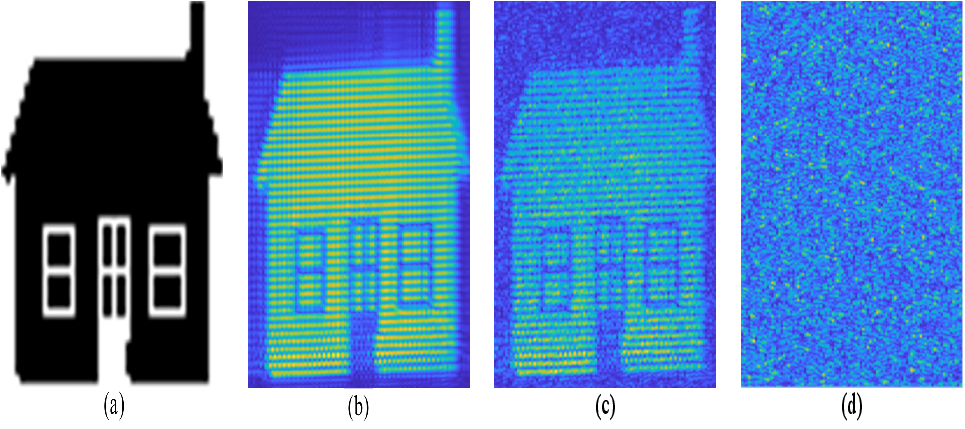}
\caption{(a) House original picture.\enspace (b) House imaging with ARIS ($a_{\rm max}=20$).\enspace (c) House imaging with ARIS ($a_{\rm max}=5$).\enspace (d) House imaging with PRIS.}
\label{fig5}\vspace{0.2 cm}
\end{figure*}

\begin{figure*}
\centering
\includegraphics[width= 15cm, height=3.75cm]{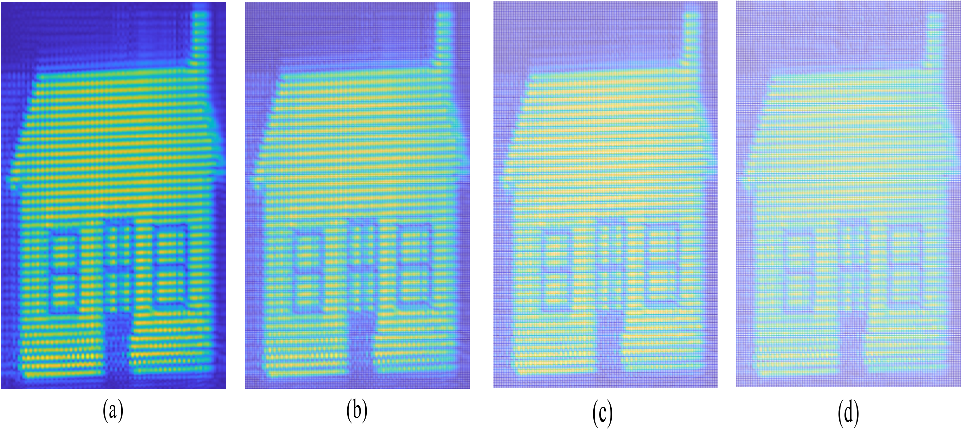}
\caption{(a) Imaging with velocity being $30m/s$.\enspace (b) Imaging with velocity being $60m/s$.\enspace (c) Imaging with velocity being $90m/s$. (d) Imaging with velocity being $120m/s$.}
\label{fig6}
\end{figure*}

The radar SNR versus the transmit power is plotted in Fig. \ref{fig4}. It can be found that the SNR rises as the transmit power increases. What is worthwhile noticing is that optimized ARIS with $a_{\rm max}=20$ scheme experiences an upward trend first, then it tends to be flat. This phenomenon occurs because the fixed $P_{\rm ARIS}$ restricts the radar performance.

Then, the SAR imaging of our proposed algorithm is presented in Fig. \ref{fig5}. We can achieve high-quality SAR imaging by utilizing a mobile ARIS with a higher amplification factor.
In contrast, the image observed with PRIS is totally blurry since the received echo signal is too weak due to the double fading effect of PRIS.

Finally, we investigate the effectiveness of our proposed algorithm under different UAV speeds in Fig. \ref{fig6}. It can be observed that the slower the UAV moves, the better the image quality. This is because the slower UAV  can provide more echo signals reflected from the imaging area.

\section{Conclusion}
In this paper, we investigated a novel  ARIS-assisted SAR imaging approach for a stationary radar system. To obtain a satisfactory imaging result of such an ARIS-assisted SAR system, we first modified the conventional RD imaging algorithm by which the stationary radar station can utilize the mobile ARIS to form a synthetic aperture for SAR imaging. Then, to achieve a better quality of  SAR imaging, we proposed an ARIS design algorithm to maximize the SNR of the echo signal at the radar receiver. Simulation results illustrated the effectiveness of ARIS-assisted imaging and our proposed algorithm. In future work, we will explore the robustness of our algorithm against UAV-induced vibrations during flight, which may cause deviations that disrupt optimization and estimation, to ensure reliable performance in real-world scenarios.


\begin{thebibliography}{00}

\bibitem{Curlander Wiley 1991} J. C. Curlander and R. N. McDonough, \textit{Synthetic aperture radar: Systems and signal processing}. New York: Wiley, 1991.

\bibitem{Cumming Artech House 2005} I. G. Cumming and F. H. Wong, \textit{Digital processing of synthetic aperture radar data: Algorithms and implementation}. Norwood, MA: Artech House, 2005.

\bibitem{Moreira GRSM 2013} A. Moreira, P. Prats-Iraola, M. Younis, G. Krieger, I. Hajnsek, and K. P. Papathanassiou, ``A tutorial on synthetic aperture radar,'' \textit{IEEE Geosci. Remote Sens. Mag.}, vol. 1, no. 1, pp. 6-43, Mar. 2013.




\bibitem{Wu C 2020} Q. Wu and R. Zhang, ``Towards smart and reconfigurable environment: Intelligent reflecting surface aided wireless network,'' \textit{IEEE Commun. Mag.},  vol. 58, no. 1, pp. 106-112, Jan. 2020.

\bibitem{Wu TWC 2019} Q. Wu and R. Zhang, ``Intelligent reflecting surface enhanced wireless network via joint active and passive beamforming,'' \textit{IEEE Trans. Wireless Commun.}, vol. 18, no. 11, pp. 5394-5409, Nov. 2019.

\bibitem{Liu CST 2021} Y. Liu, X. Liu, X. Mu, T. Hou, J. Xu, M. Renzo, and N. Dhahir ``Reconfigurable intelligent surfaces: Principles and opportunities,'' \textit{IEEE Commun. Surveys Tuts.}, vol. 23, no. 3, pp. 1546-1577, Jul. 2021.

\bibitem{Pan JSTSP 2022} C. Pan \textit{et al.}, ``An overview of signal processing techniques for RIS/IRS-aided wireless systems,'' \textit{IEEE J. Sel. Topics Signal Process.}, vol. 16, no. 5, pp. 883-917, Aug. 2022.

\bibitem{Luo TVT 2023} H. Luo, R. Liu, M. Li, and Q. Liu, ``RIS-aided integrated sensing and communication: Joint beamforming and reflection design,'' \textit{IEEE Trans. Veh. Technol.}, vol. 72, no. 7, pp. 9626-9630, Jul. 2023.


\bibitem{Shao JSAC 2022} X. Shao, C. You, W. Ma, X. Chen, and R. Zhang, ``Target sensing with intelligent reflecting surface: Architecture and performance,'' \textit{IEEE J. Sel. Areas Commun.}, vol. 40, no. 7, pp. 2070-2084, Jul. 2022.
\bibitem{Lu SL 2021} W. Lu, Q. Lin, N. Song, Q. Fang, X. Hua, and B. Deng, ``Target detection in intelligent reflecting surface aided distributed MIMO radar systems,'' \textit{IEEE Sensors Lett.}, vol. 5, no. 3, pp. 1-4, Mar. 2021.
\bibitem{Zhang TWC 2021} H. Zhang, B. Di, K. Bian, Z. Han, and L. Song, ``MetaLocalization: Reconfigurable intelligent surface aided multi-user wireless indoor localization,'' \textit{IEEE Trans. Wireless Commun.}, vol. 20, no. 12, pp. 7743–7757, Dec. 2021.
\bibitem{Hu SAC 2020} J. Hu \textit{et al.}, ``Reconfigurable intelligent surface based RF sensing: Design, optimization, and implementation,'' \textit{IEEE J. Sel. Areas Commun.}, vol. 38, no. 11, pp. 2700-2716, Nov. 2020.
\bibitem{Liu WC 2023} R. Liu, M. Li, H. Luo, Q. Liu, and A. L. Swindlehurst, ``Integrated sensing and communication with reconfigurable intelligent surfaces: Opportunities, applications, and future directions,'' \textit{IEEE Wireless Commun.}, vol. 30, no. 1, pp. 50-57, Feb. 2023.
\bibitem{Liu JSTSP 2022} R. Liu, M. Li, Y. Liu, Q. Wu, and Q. Liu, ``Joint transmit waveform and passive beamforming design for RIS-aided DFRC systems,'' \textit{IEEE J. Sel. Topics Signal Process.}, vol. 16, no. 5, pp. 995-1010, Aug. 2022.
\bibitem{Liu TWC 2024} R. Liu, M. Li, Q. Liu and A. L. Swindlehurst, ``SNR/CRB-constrained joint beamforming and reflection designs for RIS-ISAC systems,'' \textit{IEEE Trans. Wireless Commun.}, vol. 23, no. 7, pp. 7456-7470, Jul. 2024.
\bibitem{Esmaeilbeig SPL 2022} Z. Esmaeilbeig, K. V. Mishra, A. Eamaz, M. Soltanalian, ``Cramér–Rao lower bound optimization for hidden moving target sensing via multi-IRS-aided radar,'' \textit{IEEE Signal Process. Lett.}, vol. 29, pp. 2422-2426, Nov. 2022.



\bibitem{Zhang TC 2023} Z. Zhang \textit{et al.}, ``Active RIS vs. passive RIS: Which will prevail in 6G?'' \textit{IEEE Trans. Commun.}, vol. 71, no. 3, pp. 1707-1725, Mar. 2023.

\bibitem{Long TWC 2021}  R. Long, Y.-C. Liang, Y. Pei, and E. G. Larsson, ``Active reconfigurable intelligent surface-aided wireless communications,'' \textit{IEEE Trans. Wireless Commun.}, vol. 20, no. 8, pp. 4962-4975, Aug. 2021.

\bibitem{Zhu TCOM 2022}  Q. Zhu, M. Li, R. Liu, Y. Liu, and Q. Liu, ``Joint beamforming designs for active reconfigurable intelligent surface: A sub-connected array architecture,'' \textit{IEEE Trans. Commun.}, vol. 70, no. 11, pp. 7628-7643, Nov. 2022.

\bibitem{Zhu TVT 2023} Q. Zhu, M. Li, R. Liu, and Q. Liu, ``Joint transceiver beamforming and reflecting design for active RIS-aided ISAC systems,'' \textit{IEEE Trans. Veh. Technol.}, vol. 72, no. 7, pp. 9636-9640, Jul. 2023.

\bibitem{Zhu TWC 2024} Q. Zhu, M. Li, R. Liu, and Q. Liu, ``Cramer-Rao bound optimization for active RIS-empowered ISAC systems,''  \textit{IEEE Trans. Wireless Commun.}, early access, Apr. 10, 2024, doi: 10.1109/TWC.2024.3384501.

\bibitem{Rihan SPL 2022} M. Rihan, E. Grossi, L. Venturino, and S. Buzzi, ``Spatial diversity in radar detection via active reconfigurable intelligent Surfaces,'' \textit{IEEE Signal Process. Lett.}, vol. 29, pp. 1242-1246, May 2022.




\bibitem{Shen TSP 2018} K. Shen and W. Yu, ``Fractional programming for communication systems—Part I: Power control and beamforming,'' \textit{IEEE Trans. Signal Process.}, vol. 66, no. 10, pp. 2616-2630, May 2018.


\bibitem{Sun TSP 2017}Y. Sun, P. Babu, and D. P. Palomar, ``Majorization-minimization algorithms in signal processing, communications, and machine learning,'' \textit{IEEE Trans. Signal Process.}, vol. 65, no. 3, pp. 794-816, Feb. 2017.


\end{thebibliography}
\end{document}